\documentclass{egpubl}
\usepackage{eg2026}
 
\ConferencePaper        
\CGFStandardLicense

\usepackage[T1]{fontenc}

\usepackage{cite}  
\BibtexOrBiblatex
\electronicVersion
\PrintedOrElectronic
\ifpdf \usepackage[pdftex]{graphicx} \pdfcompresslevel=9
\else \usepackage[dvips]{graphicx} \fi

\usepackage{lineno}
\usepackage{xcolor}
\usepackage{amsmath} 
\usepackage{amssymb}
\usepackage{amsfonts}
\definecolor{kleinblue}{RGB}{0,47,167}

\usepackage{booktabs}

\usepackage{graphicx}
\usepackage{booktabs}
\usepackage{threeparttable}
\definecolor{bestcolor}{gray}{0.9}
\usepackage{colortbl}
\usepackage{booktabs}
\usepackage{multirow}
\usepackage{siunitx} 
\usepackage{placeins}

\title[LeafFit: Plant Assets Creation from 3D Gaussian Splatting]%
      {LeafFit: Plant Assets Creation from 3D Gaussian Splatting}

\author[C. Luo \& N. Umetani] 
{\parbox{\textwidth}{\centering 
        Chang Luo\orcid{0009-0006-2717-4598} 
        and Nobuyuki Umetani\orcid{0000-0003-1251-970X} 
      }
      \\
{\parbox{\textwidth}{\centering 
        The University of Tokyo, Japan
       }
}
}

\begin{document}

\teaser{
   \includegraphics[width=177mm]{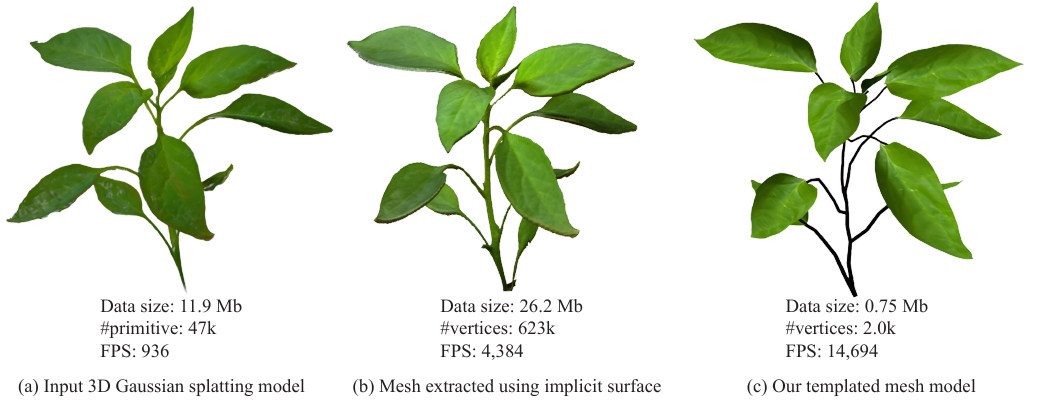}
   \centering
    \caption{Leveraging the repetition of leaf shapes, LeafFit reuses a single template leave across all the leaves in a plant. From an input 3D Gaussian splatting (3DGS) capture (a) to produce single-sided, instanced meshes that preserve thin-leaf detail (c) while drastically reducing storage and boosting frame per second. Our mesh is more compact than the mesh from implicit surface conversion of input 3DGS (b). The results are game-ready, fully editable assets, where textures and per-leaf geometry can be calculated in real time. }
  \label{fig:teaser}
}

\maketitle
\begin{abstract}
We propose LeafFit, a pipeline that converts 3D Gaussian Splatting (3DGS) of individual plants into editable, instanced mesh assets. While 3DGS faithfully captures complex foliage, its high memory footprint and lack of mesh topology make it incompatible with traditional game production workflows. We address this by leveraging the repetition of leaf shapes; our method segments leaves from the unstructured 3DGS, with optional user interaction included as a fallback. A representative leaf group is selected and converted into a thin, sharp mesh to serve as a template; this template is then fitted to all other leaves via differentiable Moving Least Squares (MLS) deformation. At runtime, the deformation is evaluated efficiently on-the-fly using a vertex shader to minimize storage requirements. Experiments demonstrate that LeafFit achieves higher segmentation quality and deformation accuracy than recent baselines while significantly reducing data size and enabling parameter-level editing. Our source code is publicly available at \href{https://github.com/netbeifeng/leaf_fit}{https://github.com/netbeifeng/leaf\_fit}.

\ccsdesc[300]{Computing methodologies~Shape modeling}
\ccsdesc[300]{Computing methodologies~Graphics systems and interfaces}

\printccsdesc   
\end{abstract}  
\section{Introduction}
Plants are ubiquitous and diverse in natural environments.
However, their assets in virtual scenes are still largely created by hand, which is a slow and labor-intensive process that requires significant expertise. 
In particular, the rich morphological variations, such as stem height and leaf sizes, make replicating each plant manually difficult to handle.
%
3D Gaussian Splatting (3DGS)~\cite{kerbl3Dgaussians} offers high-fidelity capture with efficient rendering, and it is particularly effective for vegetation with thin leaves and intricate silhouettes. 
Despite their visual quality, Gaussian reconstructions of plants see limited practical adoption in asset production. 
First, representing complex leaf structures faithfully using 3DGS requires large memory storage. 
This size challenge can be significantly mitigated by exploiting the repetition of leaf instances within an individual plant.
However, similar leaves within the same reconstruction are neither identified nor reused.
Secondly, production workflows and game engines are typically optimized for meshes and textures. Therefore, raw 3DGS primitives cannot natively benefit from standard hardware acceleration pipelines.
Conversion of Gaussian splatting into mesh with texture is possible using implicit surfaces~\cite{guedon2023sugar,Huang2DGS2024,Yu2024GOF}. 
However, these methods using implicit surfaces are not suitable for thin structures, the implicit representation tends to thicken thin sheets, cannot represent the front and back sides of the mesh with different textures, and produces overly dense geometry (see Fig.~\ref{fig:teaser}-b). 
%
This difficulty motivates a representation that factors a plant into leaf instances, aligns them non-rigidly to a shared template, and extracts a lightweight surface that preserves per-leaf variation while avoiding redundancy.
Instance-retrieval approaches that ground to external models can segment repeated objects~\cite{violante2025splatandreplace}. Still, they are effective mainly when instances are nearly identical and do not account for subtle geometric differences that commonly appear across leaves of the same species.

To address these limitations, we propose a pipeline that factors plants into leaf instances and reuses a single explicit template. 
Our system comprises three components: (i) automatic and manual segmentation of Gaussian splatting of a plant into leaf instances; (ii) non-rigid registration of leaves to preserve their shape variation; and (iii) extraction of a thin, lightweight template mesh and texture from Gaussian primitives. 
Technically, we present a robust leaf segmentation method based on the geodesic distances computed on the Gaussian primitives. 
This approach is fast and robust across various plant specimens, as it does not require a training dataset.  
Furthermore, we propose a differentiable moving least squares method that allows fitting a 3D Gaussian splatting of a leaf to another.

We evaluate our approach qualitatively by comparing leaf instance overlays and renderings against manual annotations and recent Gaussian-to-mesh baselines.
Quantitatively, our method improves segmentation accuracy and deformation fidelity while producing lighter meshes that better preserve thin structures, and it supports real-time online evaluation from compact per-leaf parameters.
Our technical contributions include:
\begin{itemize}
  \item \textbf{Instance-aware leaf segmentation:} a robust workflow that separates leaves into instances, combining automatic geodesic-based segmentation.
  \item \textbf{Template-driven non-rigid alignment:} a differentiable MLS formulation that registers each leaf to a shared template; per-leaf alignment parameters are optimized and stored once, then evaluated online to generate deformed geometry at render time.
  \item \textbf{Lightweight surface extraction and appearance transfer:} a template mesh tailored to thin leaves via point-based reconstruction with seamless texture mapping, whose geometry and appearance are reused across leaves to avoid redundancy while preserving variation.
\end{itemize}
\section{Related Works}


\begin{figure}
    \centering
    \includegraphics[width=84mm]{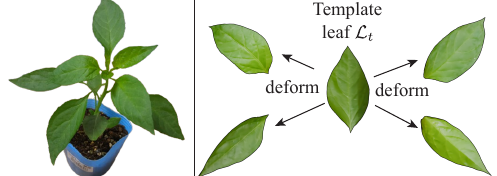}
    \caption{No two leaves belongs to a plant are exactly the same, but they are similar. We compress the data by replacing the leaves by deformed template leaf.}
    \label{fig:related_works_similar_leaves}
\end{figure}

\subsection{Game Asset Creation for Plants}
Game asset creation for vegetation has traditionally relied on manual modeling and sculpting in digital content creation tools (e.g., Maya, Blender~\cite{maya, blender}), which is highly labor-intensive and demands substantial artistic expertise. 
To reduce this burden, computer-aided approaches such as procedural methods~\cite{prusinkiewicz2012algorithmic, ijiri2006sketch, stava2014inverse, guo2018realistic, talton2011metropolis} and dedicated toolchains~\cite{ORCASpeedTree, pradal2009plantgl} have been widely developed and adopted. 
Notably, Lee et al.~\cite{lee2023latent} leverage a Transformer-based architecture for controllable tree generation, whereas template-based approaches~\cite{zhou2025treestructor} reconstruct forest scenes by retrieving and ranking models from a database.
Although these methods can automatically generate lightweight, well-structured polygon meshes that are suitable for real-time rendering and can be seamlessly integrated into game engines, most of these approaches primarily focus on generating branch structures, while leaves are often only randomly scattered across the skeleton.
To compactly represent plant's leaves, parametric templates using Bézier curves are presented ~\cite{bradley2013image, chaurasia2017editable}.
However, handling deformations between two Bézier-parametrized leaves remains challenging, as the direct deformation is often infeasible due to mismatches in control point configurations, making it difficult to establish a valid mapping without resampling.
More recently, \cite{yang2025neuraleaf} leverages neural networks to learn latent embeddings of leaf shape, texture, and deformation, enabling high-quality reconstructions but requiring training on large-scale leaf datasets.
In contrast, our work avoids the need for training feedforward networks on massive datasets.

%
%

\subsection{3D Representations for Plant Phenotyping}
A wide range of 3D representations have been explored for plant reconstruction, particularly in the context of plant phenotyping and digital twin creation~\cite{paproki2012novel, wang2013method, paulus2019measuring}. 
For comprehensive surveys, we refer readers to~\cite{okura20223d} for general 3D plant modeling and reconstruction, and to~\cite{SONG2025296} specifically for point cloud segmentation and organ identification.
Explicit methods such as meshes and point clouds are widely adopted in industry, where dense reconstructions can be obtained via Structure-from-Motion (SfM)~\cite{schoenberger2016sfm} and Multi-View Stereo (MVS)~\cite{schoenberger2016mvs}. 
While classic pipelines can tackle with large-scale reconstructions, they often suffer from geometric noise, require careful multi-camera setups, and struggle with thin structures such as leaves~\cite{nguyen2016comparison,andujar2018three,lu2020reconstruction}. 
Manual modeling, in contrast, achieves high-quality results but is extremely labor-intensive~\cite{pradal2009plantgl, ORCASpeedTree}. More recently, implicit neural methods such as Neural Radiance Fields (NeRF)~\cite{mildenhall2020nerf} have shown impressive capabilities in synthesizing photorealistic views by representing scenes as continuous volumetric functions~\cite{hu2024high, li2025plantsurvey, fruitnerf2024}. 
However, NeRFs are slow to train and memory-intensive, the implicit nature makes it difficult to extract explicit structures or support fine-grained editing~\cite{haque2023instruct}. Accelerations such as hash grid~\cite{mueller2022instant} improve performance but do not fundamentally resolve these limitations. 

Most recently, 3D Gaussian splatting (3DGS)~\cite{kerbl3Dgaussians} has emerged as a breakthrough representation that combines the visual fidelity of implicit methods with the efficiency and explicitness required for downstream use. 
By rasterizing anisotropic Gaussian primitives with learned color, opacity, and covariance parameters, 3DGS achieves real-time rendering quality that was previously out of reach. 
Building on its success, subsequent works have extended Gaussian kernel with more diverse kernel formulations~\cite{Huang2DGS2024,huang2024DRK}, enabled direct editing~\cite{wang2024gaussianeditor}, repetition localization and replacement~\cite{violante2025splatandreplace}, and explored integration with game asset generation pipelines~\cite{li2024procgs}. 

Unlike implicit neural fields, the discrete and spatially localized nature of Gaussian splatting makes them inherently well-suited for selection, manipulation, and user interaction. These properties are particularly advantageous for plant modeling, where fine-scale, repetitive structures need to be accurately extracted, aligned, and instanced~\cite{ojo2024splanting, shen2025plantgaussian}. 
In this work, we adopt Gaussian Splatting as the foundation for representing plants due to its efficiency, editability, and compatibility with both real capture pipelines and lightweight game asset workflows.

\begin{figure*}[t!]
   \centering   \includegraphics[width=176mm]{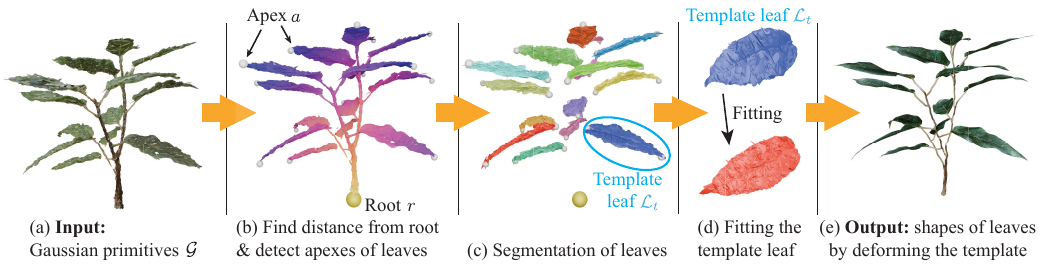}
   \caption{\textbf{Workflow.}
(a) Input: a pre-trained set of Gaussian primitives $\mathcal{G}$.
(b) The user selects a root primitive $r$; we compute geodesic distances from $r$, detect apexes $a$ as local maxima, and build a tree. The tree graph is used to find leaf petiole.
(c) Using petiole and apex cues, we segment leaves and let the user choose a template leaf $\mathcal{L}_t$.
(d) Other leaves are fitted to $\mathcal{L}_t$ via Moving Least Squares deformation.
(e) Output: an efficient, editable instanced mesh suitable for game assets.}
   \label{fig:workflow}
 \end{figure*}

\subsection{Interactive Editing of Gaussian Representations}

For editing or interacting with Gaussian-reconstructed scenes, selection and segmentation form the foundation. 
A number of recent efforts~\cite{wang2024gaussianeditor, yan20243dsceneeditor, xiao2025localized} explore scene editing guided by user prompts, where text input is used to assist object selection or deletion within the scene. 
Similarly, \cite{jaingaussiancut} employs sparse point clicks or sketches in combination with a graph-cut construction over Gaussians to achieve interactive selection. 
In contrast, \cite{playcanvas2025supersplat} provides more concrete manual selection tools, offering screen-space pickers such as lasso and brush, as well as 3D pickers such as sphere and box selection, enabling accurate control. 
However, these approaches can be time-consuming, as users often need to rotate the scene from multiple angles to complete the selection. 
None of the above-mentioned methods are specialized for plant leaf segmentation. 
In our work, we introduce an automatic segmentation algorithm based on the heat distance  method~\cite{Crane:2017:HMD}, leveraging the strong prior of leaf shape. 
We also provide a fallback interactive manual segmentation tool, also guided by the heat distance method, to handle challenging cases where automatic segmentation may fail.

Once Gaussian primitives or object parts have been reliably selected, the next step in interactive editing is to deform or align them to achieve the desired geometry. 
Several efforts explore Gaussian deformation through cage-based control~\cite{huang2024gsdeformer}, sketch input~\cite{xie2024sketch}, or physics-inspired constraints~\cite{jiang2024vr-gs}, but these methods mainly focus on in-place deformation. 
More recent studies investigate registration of Gaussian representations, either by directly exploiting Gaussian parameters for fast alignment~\cite{chang2024gaussreg} or by extending iterative closest point (ICP)~\cite{segal2009generalized} into SLAM systems~\cite{gsicpslam2024}, spline arc length parameterized non-rigid deformation~\cite{pandey2025painting}.
In the broader context of point-to-mesh alignment, hierarchical strategies combined with As-Rigid-As-Possible (ARAP) energy have been proposed to robustly register meshes to unstructured point clouds~\cite{bourquat2022hierarchical}.
While these works demonstrate the feasibility of Gaussian-based manipulation, our method leverages differentiable moving least squares (MLS)~\cite{schaefer2006image, zhu20073d} deformation, enabling robust inter-leaf alignment and providing a basis for mesh extraction tailored to vegetation assets.

\subsection{Surface Reconstruction from Gaussian Splatting}

Although learned Gaussians can be directly imported into a game scene, the representation is heavy and redundant due to its unstructured point-based nature. 
Despite strong visual fidelity, Gaussians remain image-space oriented for vegetated scenes. With lacking explicit topology, do not encode semantic part structure (e.g., individual leaves), and offering limited support for repetition and instancing that are common in plants. 
As a result, direct deployment as game assets is hindered by memory footprint, platform support, and the absence of mesh-level controls required in asset pipelines.

Existing pipelines~\cite{guedon2023sugar,Huang2DGS2024,Yu2024GOF} obtain meshes from Gaussians by binding them to implicit fields and applying isosurface extraction. 
Typically, Gaussian opacity or density is converted into a continuous signed distance field~\cite{park2019deepsdf}, and a surface is then extracted with marching cubes~\cite{lorensen1998marching}. 
This approach is general but often produces overly dense and thick geometry for thin sheets such as leaves, and is prone to staircasing or blobby artifacts. 
Subsequent decimation can reduce fidelity, and the outputs remain too heavy for real-time use. 
Moreover, implicit extraction is highly sensitive to query resolution, and struggles with thin structures, frequently generating inflated surfaces around the true geometry.
In contrast, we incorporate shape priors and employ the ball pivoting algorithm (BPA)~\cite{bernardini2002ball} to reconstruct thin and lightweight template leaf meshes directly from Gaussians. 
This preserves per-leaf individuality while supporting instancing, yielding game-ready assets that are both efficient and faithful to observed structures, surpassing implicit-field isosurfaces in compactness and procedural synthesis in realism.

\section{Methods}
\paragraph*{Input and Output}

The overall workflow of our method is shown in Figure~\ref{fig:workflow}. 
Let $\mathcal{G}=\{g_i\}_{i=1}^{|\mathcal{G}|}$ be an input 3D Gaussian splatting scene~\cite{kerbl3Dgaussians} of a single plant, where each Gaussian primitive $g_i$ stores a 3D center, covariance, opacity, and view-dependent color coefficients. 
Note that this scene contains only a single plant, as we trim out the background beforehand. 
Then, we segment the input Gaussian primitives $\mathcal{G}$ into per-leaf groups $\{\mathcal{L}_j\}_{j=1}^{|\mathcal{L}|}$ as described in Sec.~\ref{sec:segmentation}. 
The user chooses one group as the template $\mathcal{L}_t$.
The outputs are: (i) a template mesh $M=(V,F)$ reconstructed from $\mathcal{L}_t$, (ii) position of the control points for all the leaves $\{C_j\}_{j=1}^{|\mathcal{L}|}$. 
By moving the control points of the template leaf $C_t$ to target leaf $C_j, \{j\ne t\}$, we can define the moving least squares (MLS) deformation field $\Phi_j$ that moves the vertices of the template mesh such that the template mesh fits the shape of the leaf (see Sec.~\ref{sec:registration}).
The mesh extraction procedure is described in Sec.~\ref{sec:extraction}.
We do not store per-leaf meshes; instead, they are generated on the fly by applying the MLS deformation induced by $C_j$ to $V$ at loading time or in real-time using a vertex shader. 

%
%

\subsection{Leaf Instance Segmentation}
\label{sec:segmentation}
We denote each segmented leaf as $\mathcal{L}_j$, and the remaining Gaussian primitives that do not belong to any leaf segment are grouped as the stem $\mathcal{S}$. 
Together they form a partition of the plant Gaussians $\mathcal{G} = \{\mathcal{L}_j\} \cup \mathcal{S}$.
Segmentation is achieved by combining automatic detection based on geodesic distances with optional manual refinement.

\paragraph*{Distance from Root}
From the Gaussian splatting of a scene $\mathcal{G}$,  the user selects the Gaussian $r$ that is closest to the root. 
This root Gaussian is the source for the distance field computation using heat propagation~\cite{Crane:2017:HMD}. 
Note that we regard a Gaussian splat as a point cloud using only the center coordinates of the Gaussian primitives.
As a result, we obtain the geodesic distance defined at each Gaussian $D_r[i]$, where $D_r[r]=0$ and values increase smoothly towards the leaf apex (i.e., leaf tip).

\begin{figure}[b!]
    \centering
    \includegraphics[width=84mm]{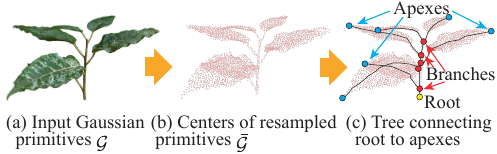}
       \caption{\textbf{Tree-graph construction.}
    (a) Start from the full set of Gaussians $\mathcal{G}$.
    (b) Downsample to a sparse subset $\bar{\mathcal{G}}$ to reduce the cost of detecting apexes via local maxima on the geodesic field (from the root $r$).
    (c) For each apex $\alpha$, trace a rootward path by greedy descent on geodesic distance; when it meets an existing path, attach it as a branch to complete the tree.}
    \label{fig:apex}
\end{figure}

\paragraph*{Leaf Apexes}
To segment the leaf, we first detect leaf apexes by computing the local maxima of distance from the root $D_r[\cdot]$.
However, because each leaf is represented by many non-uniformly distributed Gaussian primitives, directly identifying local maxima on the dense input is prone to noise.
Hence, to stabilize the computation, we first apply farthest point sampling to sample the Gaussian primitives uniformly (see Fig.~\ref{fig:apex}).
We denote the set of sampled Gaussian primitives as $\bar{\mathcal{G}} \subset \mathcal{G}$.
The sample size $N_s = |\bar{\mathcal{G}}|$ is chosen empirically to ensure sufficient density for robust topological analysis.
Note that farthest point sampling tends to preserve points at the boundaries, effectively retaining potential leaf tips.
We further compute $N_k$ nearest neighbors~\cite{friedman1977kdtree} among the sampled Gaussian primitives $\bar{\mathcal{G}}$ using the Euclidean distance.
A sampled Gaussian $i$ is registered as a candidate apex if it is a strict local geodesic maximum, while all its $N_k$ Euclidean neighbors $j$ satisfy $D_r[j] < D_r[i]$.

%

\paragraph*{Tree Construction}
A leaf can have multiple local maxima, so a single apex does not directly correspond to a unique leaf.
To analyze the connectivity of each leaf, we construct a tree that connects the root to the apexes, where its branches consist of sampled Gaussian primitives.
We trace a path connecting the sparsely sampled Gaussians $\bar{\mathcal{G}}$ from the apex $a$ to the root $r$, denoted as $\mathcal{P}_a=\{p_k \mid p_k \in \bar{\mathcal{G}}, p_1=a, p_{|\mathcal{P}|}=r \}$.
The path follows the steepest descent of the geodesic distance $D_r$:
\begin{equation}
p_{k+1} = \arg\min_{g\in\mathcal{N}(p_k)} D_r[g].
\end{equation}
When paths from different apexes intersect, their shared Gaussian is marked as a junction of the tree.
To avoid premature joins caused by noise (e.g., paths accidentally touching at boundaries), we adopt a \textit{deferred-merge rule}.
Specifically, if the neighborhood of the current point $p_k$ contains a node labeled as visited by another path, we do not terminate immediately.
Instead, we continue one step further along the steepest descent to $p_{k+1}$; we merge the paths only if the neighborhood of $p_{k+1}$ also contains visited vertices.
This ensures that the merge occurs well within the shared branch structure rather than at a noisy interface.

%
%
%

\begin{figure}[t!]
    \centering
    \includegraphics[width=84mm]{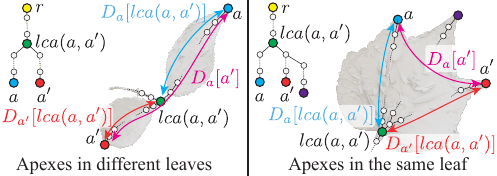}
\caption{\textbf{Separating apex pairs via triangle inequality.}
Given two apexes $a$ and $a'$ with lowest common ancestor $\mathrm{lca}(a,a')$ on the rootward tree, let $D_a[\cdot]$ and $D_{a'}[\cdot]$ denote geodesic distances from $a$ and $a'$, respectively.
\textbf{Left:} Apexes on different leaves satisfy $D_a[a'] \approx D_a[\mathrm{lca}(a,a')] + D_{a'}[\mathrm{lca}(a,a')]$.
\textbf{Right:} Apexes on the same leaf yield a strictly shorter direct path,
$D_a[a'] < D_a[\mathrm{lca}(a,a')] + D_{a'}[\mathrm{lca}(a,a')] - \tau$.
We use this criterion to group apexes into leaves.}
    \label{fig:separation}
\end{figure}

\paragraph*{Grouping Apexes of a Leaf}
For complex leaves such as maple, multiple apexes exist on a single leaf (see Fig.~\ref{fig:separation}-right).
Hence, we group apexes that belong to the same leaf. 
First, we compute the geodesic distance from each apex $D_a[\cdot]$.
Similar to the distance from root $D_r$, the distance from apex is computed for all the Gaussian primitives $\mathcal{G}$ using the heat distance method~\cite{Crane:2017:HMD}.
To check if two apexes $a$ and $a'$ belong to the same leaf, we compare the direct geodesic distance $D_a[a']$ against the path through their common tree structure.
First, we identify the lowest common ancestor (LCA) node in the tree, denoted as $\mathrm{lca}(a,a')\in\bar{\mathcal{G}}$.
According to the triangle inequality, the direct path is shorter or equal to the path via the LCA. 
Specifically, we consider two apexes to be on the \textit{same} leaf if the direct geodesic distance is strictly shorter than the tree-based path by a margin $\tau$:
\begin{equation}
    D_a[a'] < D_a[\mathrm{lca}(a,a')] + D_{a'}[\mathrm{lca}(a,a')] - \tau,
    \label{eqn:tri_ineq}
\end{equation}
where $\tau$ is a distance margin parameter.
Otherwise, if the left and right hand sides are nearly equal (within $\tau$), the path must traverse the stem, implying the apexes belong to different leaves.
%
At the end, we construct an adjacency graph where edges connect apexes satisfying Eq.~\eqref{eqn:tri_ineq}, and final leaf instances are obtained as the connected components of this graph.

\paragraph*{Leaf Segmentation}
Once leaf apexes are grouped into leaf instances, the leaf petioles can be determined.
For each leaf instance, we designate the apex with the largest geodesic distance from the root as the primary tip $a$.
To locate the leaf petiole, we traverse the pre-computed path $\mathcal{P}_a$ from the tip $a$ towards the root.
At each path point $p_k \in \mathcal{P}_a$, we estimate the local leaf diameter by analyzing a geodesic slice.
Specifically, we query the set of Gaussian primitives whose geodesic distance from the apex, $D_a[\cdot]$, falls within a narrow iso-geodesic ring of width $\delta$ centered at $D_a[p_k]$. We denote this local band $\mathcal{G}_{p_k}$ as:
\begin{equation}
\mathcal{G}_{p_k} = \left\{ g \mid \left| D_a[g] - D_a[p_k] \right| \leq \frac{\delta}{2} \right\}.
\label{delta}
\end{equation}
The local diameter $d_{p_k}$ is defined as the maximum Euclidean distance between the path node $p_k$ and the retrieved primitives in the set $\mathcal{G}_{p_k}$, formulated as: 
\begin{equation}
d_{p_k} = \max_{g \in \mathcal{G}_{p_k}} \| p_k - g \|_2.
\end{equation}
As we march rootward, the iteration over path will be terminated when the diameter $d_{p_k}$ drops below a threshold $\epsilon$, indicating the petiole.
Crucially, to prevent premature termination at narrow leaf tips (e.g., in elongated leaves), we apply an \textit{early protection window} controlled by a ratio $\rho$: base detection is disabled within the initial fraction $\rho$ of the path segment from $a$ to the first branching junction in the tree graph.
All primitives with $D_a[\cdot]$ smaller than the determined base distance are assigned to the leaf.

\begin{figure}[b!]
    \centering
    \includegraphics[width=84mm]{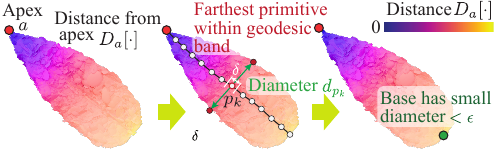}
\caption{\textbf{Leaf base determination.}
For each apex $a$, we compute the geodesic distances $D_a[\cdot]$ and traverse the rootward path $\mathcal{P}_a$. At each path point $p_k$, we query the set of primitives $\mathcal{G}_{p_k}$ within an iso-geodesic band of width $\delta$ to compute the local diameter $d_{p_k}$. The traversal terminates when $d_{p_k}$ drops below the threshold $\epsilon$, marking the petiole.}

    \label{fig:base}
\end{figure}
\begin{figure}[t!]
    \centering
    \includegraphics[width=84mm]{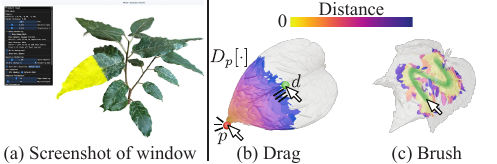}
\caption{\textbf{Manual segmentation tools.}
(a) GUI for editing Gaussian primitives; selected primitives are highlighted in yellow.
(b) \textbf{Drag selection:} Selection expands based on the geodesic distance from the apex.
(c) \textbf{Brush selection:} Primitives within a fixed geodesic radius around the cursor are selected.}

\label{fig:manual_seg}
\end{figure}

\paragraph*{Manual Segmentation}
In addition to the automatic segmentation described above, we provide an interactive tools with two selection methods, \textit{drag} and \textit{brush}, to refine the results (see Figure~\ref{fig:manual_seg}).
These tools are particularly effective for correcting segmentation in noisy regions or for handling structurally complex plants where automatic methods may fail.

For the \textit{drag tool}, the user clicks on a leaf apex and drags the cursor toward the base.
Upon the initial button press, we identify the source Gaussian $p\in\mathcal{G}$ via ray-casting. This point serves as the source for a real-time geodesic distance field computation $D_p[\cdot]$.
As the mouse is dragged, we continuously update the target Gaussian $d\in\mathcal{G}$ under the cursor.
Any Gaussian primitive $g$ is highlighted and selected if it lies within the geodesic radius defined by the current cursor position, i.e., $D_p[g] \le D_p[d]$.

On the other hand, by using the \textit{brush tool}, users select Gaussian primitives within an adjustable geodesic radius $r$ centered at the Gaussian $d$ under the cursor.
Specifically, we compute the heat geodesic distance from $d$ and select any primitive $g$ satisfying $D_d[g] \le r$.
To ensure real-time performance for these ray-Gaussian intersections, we utilize a Bounding Volume Hierarchy (BVH) structure constructed on the Gaussian primitives.
Once all leaf instances are segmented, the remaining unassigned primitives are grouped as the stem segment $\mathcal{S}$.

\subsection{Leaf Registration using Moving Least Squares}
\label{sec:registration}
Once all the leaves $\{\mathcal{L}_j\}$ are segmented, we first denoise the leaf point clouds using a moving least squares (MLS) projection.
Specifically, we locally fit a plane to the center of the Gaussian primitives and remove outliers that lie far from this surface~\cite{amenta2004defining}.
This filtering yields a clean leaf point set for stable rendering and downstream alignment. 
The user then selects a template leaf $\mathcal{L}_t$ to serve as the template leaf for alignment.

\paragraph*{Rasterization of Leaves}
We employ the 2D rasterizer from the Gaussian splatting pipeline~\cite{kerbl3Dgaussians} to render each segmented leaf and extract its depth and color in image space.
To maximize projection quality, each leaf is first PCA-aligned so that its principal plane coincides with the $xy$-plane and its normal points along the positive $z$-axis. 
This preprocessing improves both the depth computation for registration and the subsequent texture sampling of the template mesh.

We use farthest point sampling on each leaf's Gaussian primitives $\mathcal{L}_j$ to generate a sparse set of control points $C_j\in\mathbb{R}^{K\times 3}$, where $K$ is the sampled control points amount per leaf.
%
Given the two sampled control points set $C_t \subset \mathcal{L}_t$ (for template leaf) and $C_j \subset \mathcal{L}_j$ (for target leaf), 
we need firstly establish a one-to-one correspondence between their control points $C_t$ and $C_j$, as an initial guessed correspondence is required for starting the optimization of MLS deformation.
This initial correspondence assignment is solved by minimizing the global correspondence cost using the Jonker-Volgenant algorithm~\cite{crouse2016implementing}.

From this initialization, we optimize the target control point positions $C_j$ to fit the template leaf to the target leaf.
The MLS deformation field $\Phi_j: \mathbb{R}^3\rightarrow\mathbb{R}^3$ is parameterized by the source $C_t$ and target $C_j$ control handles.
First, we compare the depth images of the deformed template and the target leaf rendered from the PCA-aligned view. 
Let $D_{\Phi_j}$ and $D_{\mathcal{L}_j}$ denote the depth maps of $\Phi_j(\mathcal{L}_t)$ and $\mathcal{L}_j$, respectively. 
The depth loss is defined as:
\begin{equation}
L_{\text{depth}} = \frac{1}{|\Omega|}\sum_{u\in\Omega} \left| D_{\Phi_j}(u)-D_{\mathcal{L}_j}(u)\right|^2,
\end{equation}
where $\Omega$ is the set of foreground pixels. Second, we regularize the alignment in 3D space using a bidirectional Chamfer distance:
\begin{equation}
\begin{split}
L_{\text{chamfer}} = & \frac{1}{|\mathcal{L}_t|}\sum_{p\in \mathcal{L}_t}\min_{q\in \mathcal{L}_j}\lVert \Phi_j(p)-q\rVert^2 \\
& + \frac{1}{|\mathcal{L}_j|}\sum_{q\in \mathcal{L}_j}\min_{p\in \mathcal{L}_t}\lVert \Phi_j(p)-q\rVert^2.
\end{split}
\end{equation}
Finally, we solve for the optimal control points $C_j$ by minimizing the composite objective:
\begin{equation}
\min_{C_j} \;\; \,L_{\text{depth}}(C_j) \;+\; \lambda L_{\text{chamfer}}(C_j),
\end{equation}
where $L_{\text{depth}}$ enforces depth-map consistency in the PCA-aligned view and $L_{\text{chamfer}}$ promotes 3D geometric agreement.

\paragraph*{GPU instancing and memory footprint.}
At runtime, we store only the template mesh $(V, F)$, its texture, and per-leaf kernel data of size $K\times N$. 
MLS deformation is evaluated on the fly in a vertex shader, which writes deformed positions to a GPU buffer consumed by the vertex stage. 
This keeps CPU cost negligible and avoids duplicating geometry; memory scales as $O(|V|)$ shared $+\,O(KN)$ per plant, while the per-leaf compute is $O(|V|K)$ and parallelizable.

\subsection{Template Leaf Mesh Extraction}\label{sec:extraction}
The deformation computed in the Gaussian domain can be directly reused to deform mesh vertices, enabling a consistent representation across both modalities. 
Since Gaussian splats are not yet widely supported in many content creation pipelines and game engines, we extend our approach to produce explicit meshes.
Only the template leaf needs to be meshed; all other leaves are then obtained by applying the previously estimated MLS deformations to this template.

\paragraph*{Surface Topology Reconstruction}
Because leaves have thin, sheet-like structures, implicit isosurface extraction methods often fail to capture their geometry reliably, tending to produce inflated artifacts.
Instead, we downsample the template leaf Gaussians using farthest point sampling to obtain a representative point cloud, and reconstruct a watertight surface with the ball pivoting algorithm (BPA)~\cite{bernardini2002ball}.
While BPA is known at recovering thin surfaces, it is sensitive to the non-uniform density of Gaussian primitives, which can result in small topological gaps.
We therefore employ post-processing to repair and fill holes~\cite{holes_in_mesh}, yielding a clean triangular mesh $M=(V, F)$ suitable for further deformation.

\paragraph*{Texture Extraction}
For texture computation, we first align the leaf using PCA, similar to the depth image computation.
Then, we render the leaf from the back and front sides.
To support standard game engine rendering, each original triangle is duplicated into a front–back pair with separate UV patches.
Vertices are also duplicated per side with split normals to prevent UV conflicts. This construction ensures spatial consistency between the Gaussian appearance model and the extracted mesh textures, allowing for seamless transfer of visual detail from 3D Gaussian splatting to conventional mesh-based pipelines.

\paragraph*{Stem Generation}
As mentioned in Sec.~\ref{sec:segmentation}, we construct a tree by connecting the resampled Gaussian primitives $\bar{\mathcal{G}}$ from apexes to the root $r$.
We generate the stem geometry by replacing the edges of the tree that do not belong to the leaf with cylinders.
In our current implementation, we determine the radius of the cylinders using Leonardo's formula~\cite{Soneira1988Leonardo}, i.e., the cross-sectional area of a trunk is equal to the sum of the cross-sectional areas of its branches.
The user specifies the radius of the root and that of the petioles to ground the interpolation via GUI.
This paper focuses on the instantiation of the leaves, and a more accurate reconstruction of the stem based on Gaussian primitives is future work.

\begin{figure}[!t]
  \centering
  \includegraphics[width=0.99\linewidth]{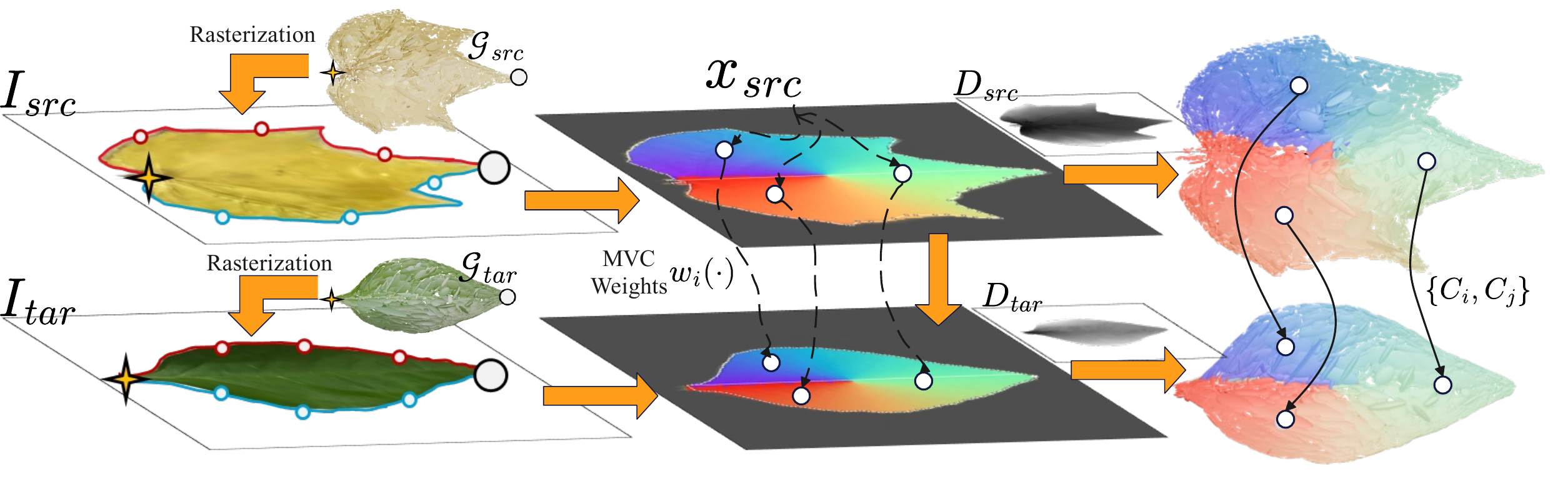}
  \caption{\textbf{Correspondence for leaf retargeting across species.}
  Orthographic renders $(I_{\mathrm{src}}, I_{\mathrm{tar}})$ with depths $(D_{\mathrm{src}}, D_{\mathrm{tar}})$ produce silhouettes that define source/target cages.
  Interior samples are transferred by MVC and back-projected with depth to form dense 3D pairs $\{(C_i, C_j)\}$ that drive MLS retargeting.}
  \label{fig:mvc_corr}
  \vspace*{-0.25cm}
\end{figure}
\subsection{Post Editing of Plant}
\label{editability}

Our scene is represented by a \emph{single} template mesh driven by a sparse set of control points via MLS deformation. 
This design decouples appearance from geometry, allowing both to be easily edited while remaining efficient.

\begin{figure}[!b]
  \centering
  \includegraphics[width=84mm]{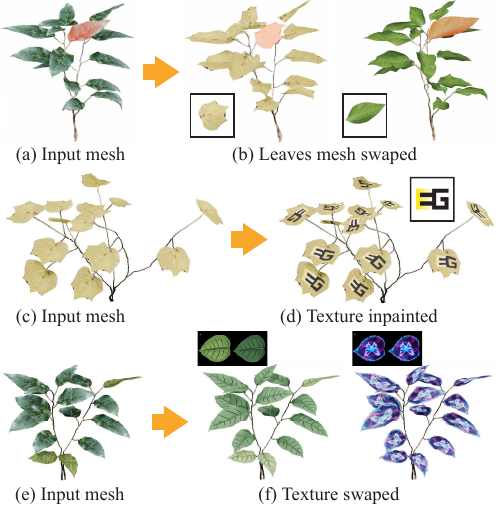}
  \caption{\textbf{Editing and retargeting.}
  Geometry retargeting from a reconstructed rubber-tree leaf to target leaves (top), and texture edits that propagate to all instances, including image inpainting overlay (middle) and full texture replacement (bottom).}
  \label{fig:editing}
\end{figure}

\paragraph*{Texture editing}
Since every leaf is a deformation of the same template mesh, we can edit the template UV texture once and propagate the change to all instances through the MLS warp. 
In practice, we apply masked inpainting or image-generation models to the template texture (standard 2D editing), which is significantly more efficient than operating directly on a dense Gaussian field. 
We demonstrate two editing workflows in Fig.~\ref{fig:editing}: (1) manual composition for input mesh (a), where users overlay external images or decals onto the texture, and (2) full texture regeneration for input mesh (c), leveraging generative AI services to completely alter the plant's appearance style.

\paragraph*{Geometry retargeting across species}
Straightforward 3D correspondence matching via farthest point sampling can be brittle for large shape gaps, so we adopt a 2D cage–transfer strategy (see Fig.~\ref{fig:mvc_corr}).
We orthographically render source and target to obtain images $I_{\mathrm{src}}, I_{\mathrm{tar}}$ and depths $D_{\mathrm{src}}, D_{\mathrm{tar}}$.
Using the segmentation results, we trace the leaf silhouettes from the apex to the petiole base.
By uniformly sampling along these left and right boundaries, we construct a 2D cage surrounding each leaf.
Interior samples $x_{\mathrm{src}}$ are drawn inside the source cage (via farthest point sampling); their mean value coordinates (MVC)~\cite{mvc2003} $w_i(x_{\mathrm{src}})$ are evaluated w.r.t.\ the source cage and used to transfer them to the target cage:
\begin{equation}
x_{\mathrm{tar}}=\frac{\sum_i w_i(x_{\mathrm{src}})\,b_i^{\mathrm{tar}}}{\sum_i w_i(x_{\mathrm{src}})}.
\end{equation}
Finally, back-projecting $(x_{\mathrm{src}}, x_{\mathrm{tar}})$ with their respective depth values $(D_{\mathrm{src}}, D_{\mathrm{tar}})$ yields dense 3D correspondence pairs $\{(C_i, C_j)\}$, which we use to fit an MLS deformation for robust retargeting.

\FloatBarrier
\section{Results}
\begin{figure*}[t!]
   \centering   \includegraphics[width=170mm]{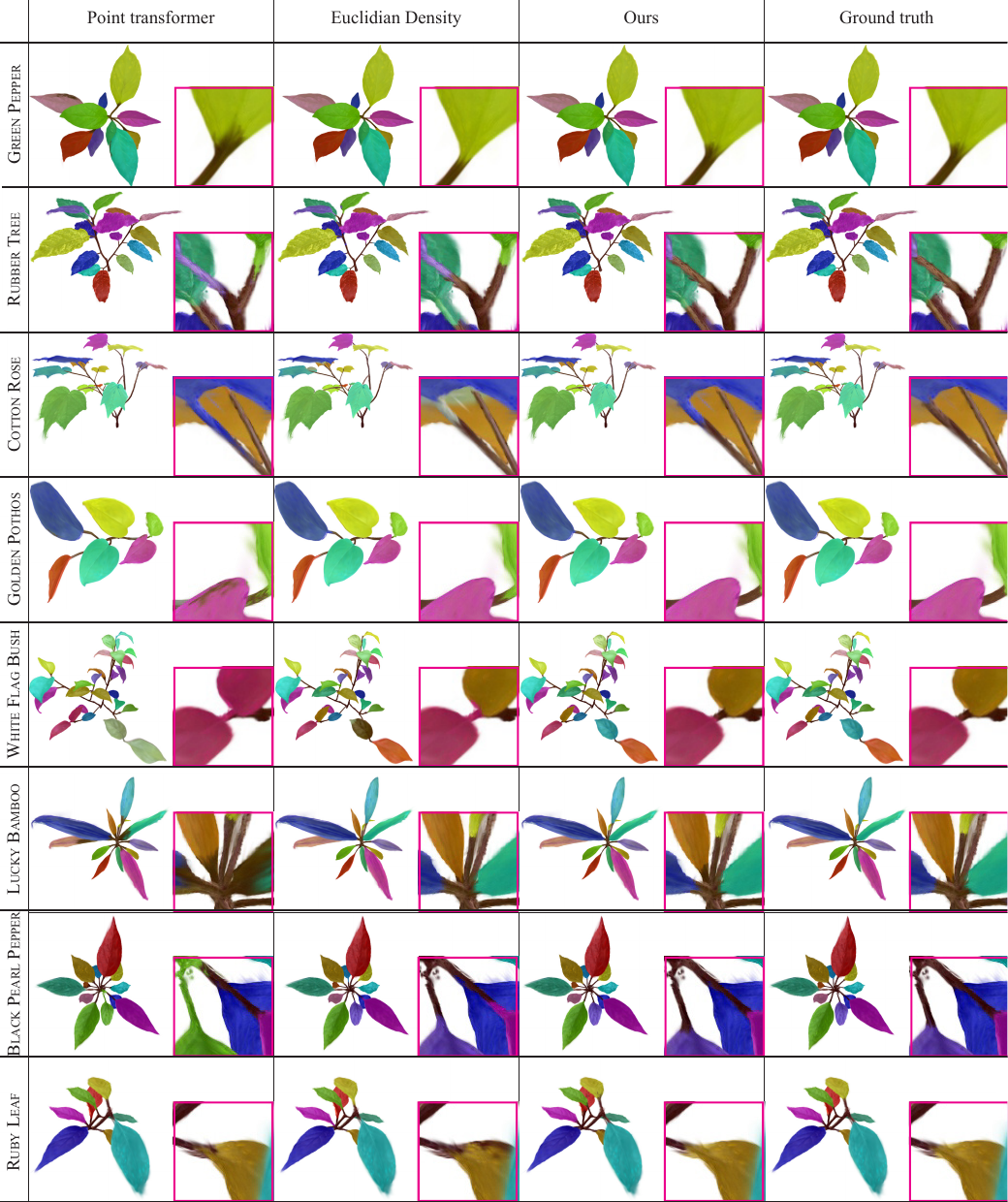}
   \caption{\textbf{Segmentation qualitative comparison.} We compare our automatic segmentation against a learned baseline (Point Transformer~\cite{wu2024ptv3}) and a Euclidean-density heuristic. Ground truth is manually annotated. Instance correspondences are matched via Hungarian assignment (IoU$\geq$0.5). Despite being training-free, our method achieves cleaner separation of overlapping leaves compared to the learned baseline, which struggles with unseen plant structures.}
   \label{fig:exp_segmentation}
 \end{figure*}
 \newcommand{\dec}[2]{#1 & #2}
\newcommand{\best}[1]{\textbf{#1}}
\newcommand{\bestdec}[2]{\textbf{#1} & \textbf{#2}}
\newcommand{\cmark}{\ding{51}} 
\newcommand{\xmark}{\ding{55}} 
 \begin{table*}[!t]
  \centering
  \setlength{\tabcolsep}{6pt}
  \begin{tabular}{
    @{} l|                 
    r@{.}l r@{.}l          
    r@{.}l r@{.}l          
    r@{.}l r@{.}l          
    r@{.}l r@{.}l          
    r@{.}l r@{.}l          
    c @{}                  
  }
    \toprule
      & \multicolumn{4}{c}{Acc $\uparrow$}
      & \multicolumn{4}{c}{mIoU $\uparrow$}
      & \multicolumn{4}{c}{mF1 $\uparrow$}
      & \multicolumn{4}{c}{PQ $\uparrow$}
      & \multicolumn{4}{c}{Time (s) $\downarrow$}
      & \textit{Training-free} \\
    \cmidrule(lr){2-5}\cmidrule(lr){6-9}\cmidrule(lr){10-13}\cmidrule(lr){14-17}\cmidrule(lr){18-21}
    Method
      & \multicolumn{2}{c}{Mean} & \multicolumn{2}{c}{Std}
      & \multicolumn{2}{c}{Mean} & \multicolumn{2}{c}{Std}
      & \multicolumn{2}{c}{Mean} & \multicolumn{2}{c}{Std}
      & \multicolumn{2}{c}{Mean} & \multicolumn{2}{c}{Std}
      & \multicolumn{2}{c}{Mean} & \multicolumn{2}{c}{Std}
      &  \\
    \midrule
    Point Transformer~\cite{wu2024ptv3}
      & \dec{94}{52} & \dec{4}{37}
      & \dec{86}{88} & \dec{9}{26}
      & \dec{89}{54} & \dec{8}{29}
      & \dec{89}{36} & \dec{7}{90}
      & \dec{16}{534} & \dec{10}{062}
      & $\times$ \\
    Euclidean Density
      & \dec{97}{26} & \dec{1}{92}
      & \dec{95}{16} & \dec{2}{92}
      & \dec{97}{35} & \dec{1}{75}
      & \dec{94}{96} & \dec{3}{39}
      & \bestdec{3}{786} & \bestdec{2}{334}
      & $\checkmark$ \\
    \midrule
    Ours 
      & \bestdec{98}{95} & \bestdec{0}{58}
      & \bestdec{98}{20} & \bestdec{0}{95}
      & \bestdec{99}{08} & \bestdec{0}{49}
      & \bestdec{98}{20} & \bestdec{0}{95}
      & \dec{4}{456} & \dec{2}{724}
      & $\checkmark$ \\
    \bottomrule
  \end{tabular}
 \caption{\textbf{Segmentation quantitative comparison.} Best scores are bolded.
Metrics are reported as mean\,(\(\pm\) std) over all plants.
\emph{Ours} consistently outperforms both the learned and heuristic baselines across all metrics (Acc, mIoU, mF1, PQ), demonstrating robust generalization without requiring training data.}
  \label{tab:exp_seg}
\end{table*}
\begin{figure*}[t]
  \centering
  \includegraphics[width=174mm]{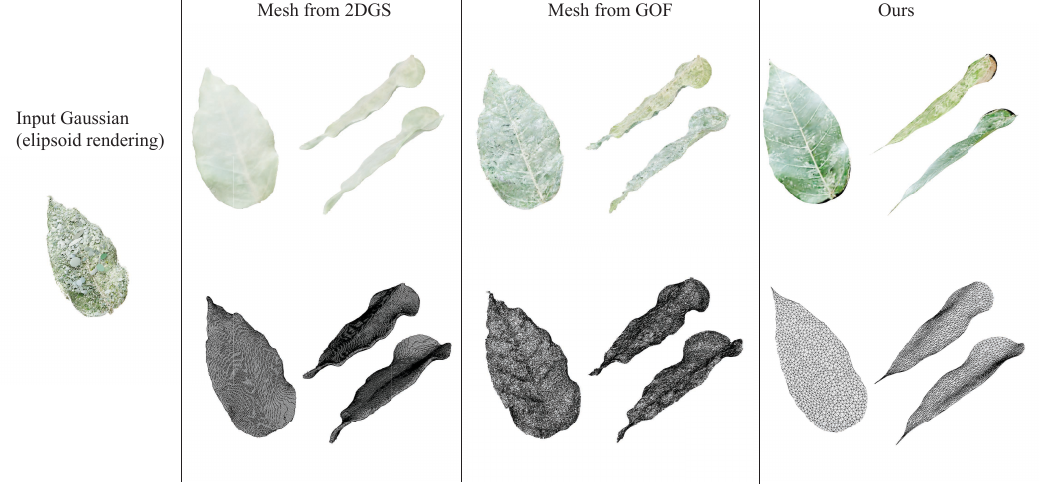}
  \caption{\textbf{Mesh extraction qualitative comparison (\textsc{Rubber Tree}).}
  Left: Input Gaussian primitives. Middle: Meshes extracted from implicit baselines (2DGS~\cite{Huang2DGS2024}, GOF~\cite{Yu2024GOF}). Right: Meshes from our conversion. Implicit methods suffer from artificial thickening (double walls) and topological noise on thin leaf structures. In contrast, our method faithfully recovers thin, sharp leaf blades with compact triangulation and clean textures.}
  \label{fig:exp_mesh}
\end{figure*}
We evaluate our approach across four axes: (i) instance segmentation, (ii) mesh quality and compactness, (iii) non-rigid deformation accuracy, and (iv) editability demonstrated via texture or geometry editing. Quantitative metrics and qualitative visualizations are provided in the following.

\paragraph*{Segmentation Evaluation}
We evaluate across diverse species and leaf morphologies. Manual annotations serve as ground truth; qualitative comparisons are in Fig.~\ref{fig:exp_segmentation} and quantitative results in Tab.~\ref{tab:exp_seg}. 
Segmentation comparisons show clean separation of overlapping leaves and suppression of stray stem regions. 
We report Accuracy (Acc), mean IoU (mIoU), mean F1 (mF1), Panoptic Quality (PQ), and wall-clock time (mean\,\(\pm\)\,std across plants). 
PQ~\cite{kirillov2019panoptic} measures joint detection+mask quality and, when the stem is a class, specifically penalizes leaf–stem confusions. 
Baselines include: (i) a \textit{training-free} Euclidean-density heuristic on a \(k\)-NN graph, which isolates the stem based on low density \(\rho\) and high gradient \(\|\nabla\rho\|\) (identifying the base at the last pre-minimum along the rootward path); and (ii) a \textit{learned} Point Transformer v3~\cite{wu2024ptv3}, trained on the PLANesT-3D dataset~\cite{mertouglu2024planest} and evaluated on our data.

\paragraph*{Mesh extraction evaluation}
We evaluate surface reconstruction for both fidelity and compactness at two scales: plant and leaf. 
In Tab.~\ref{tab:mesh_and_rendering}, we report vertex and face counts, file size, and frames per second (FPS) for real-time rendering (plant only). 
As implicit baselines, we include 2DGS~\cite{Huang2DGS2024} and GOF~\cite{Yu2024GOF}. 
Specifically, 2DGS fits depth-aware 2D Gaussians in screen space and lifts them to an implicit surface, whereas GOF directly optimizes a 3D Gaussian opacity field. In both cases, the resulting fields are converted to meshes via Marching Cubes~\cite{lorensen1998marching} for comparison.
We compare their rendered quality, model size, and extracted meshes to assess representation accuracy, rendering speed, and storage requirements. 
As summarized in the table, the plant-level mesh attains the highest FPS with the smallest storage footprint, while the leaf-level mesh maintains a fixed 2{,}048-vertex budget and the smallest file size (per-leaf FPS is not applicable). 
Qualitatively (Fig.~\ref{fig:exp_mesh}, \textsc{Rubber Tree}), the extracted mesh preserves thin blades and sharp tips, recovers clear, high-frequency texture from Gaussian splatting, and exhibits uniformly distributed vertices with well-shaped triangles. 
Compared with implicit baselines, our thin template surface reveals reduced thickening and clean sharpness along edges, while maintaining low triangle counts.

\begin{table}[t]
  \centering
  \small

  \sisetup{group-separator={,}}
  \setlength{\tabcolsep}{4pt} 
  \begin{tabular}{@{} l l | r r r @{}}
    \toprule
    Method & Representation & FPS $\uparrow$ & Size $\downarrow$ & Verts $\downarrow$ \\
    \midrule
    \multirow{3}{*}{2DGS~\cite{Huang2DGS2024}} & Gaussian (Plant) & 653.93 & 19.08 & \num{76943} \\
    & Mesh (Plant)     & 6,741.25 & 38.40 & \num{741502} \\
    & Mesh (Leaf)  & \multicolumn{1}{c}{---} & 7.26 & \num{218245} \\
    \midrule
    \multirow{3}{*}{GOF~\cite{Yu2024GOF}} & Gaussian (Plant) & 375.80 & 20.76 & \num{77996} \\
    & Mesh (Plant)     & 2,934.32 & 46.69 & \num{1107809} \\
    & Mesh (Leaf)  & \multicolumn{1}{c}{---} & 9.90 & \num{253576} \\
    \midrule
    \multirow{2}{*}{Ours} & Mesh (Plant)     & \best{11,980.11} & \best{1.13} & \best{2432} \\
    & Mesh (Leaf)  & \multicolumn{1}{c}{---} & \best{0.334} & \best{2048} \\
    \bottomrule
  \end{tabular}
\caption{\textbf{Rendering and storage efficiency.} Best scores are bolded.
Plant-level rows report FPS, storage size (MB), and vertex count.
Our explicit mesh representation achieves superior rendering speed and compression rates compared to implicit baselines (2DGS, GOF), utilizing a minimal vertex budget (sum of template vertices and control points).}
  \label{tab:mesh_and_rendering}
\end{table}

\begin{figure*}[t!]
  \centering
  \includegraphics[width=178mm]{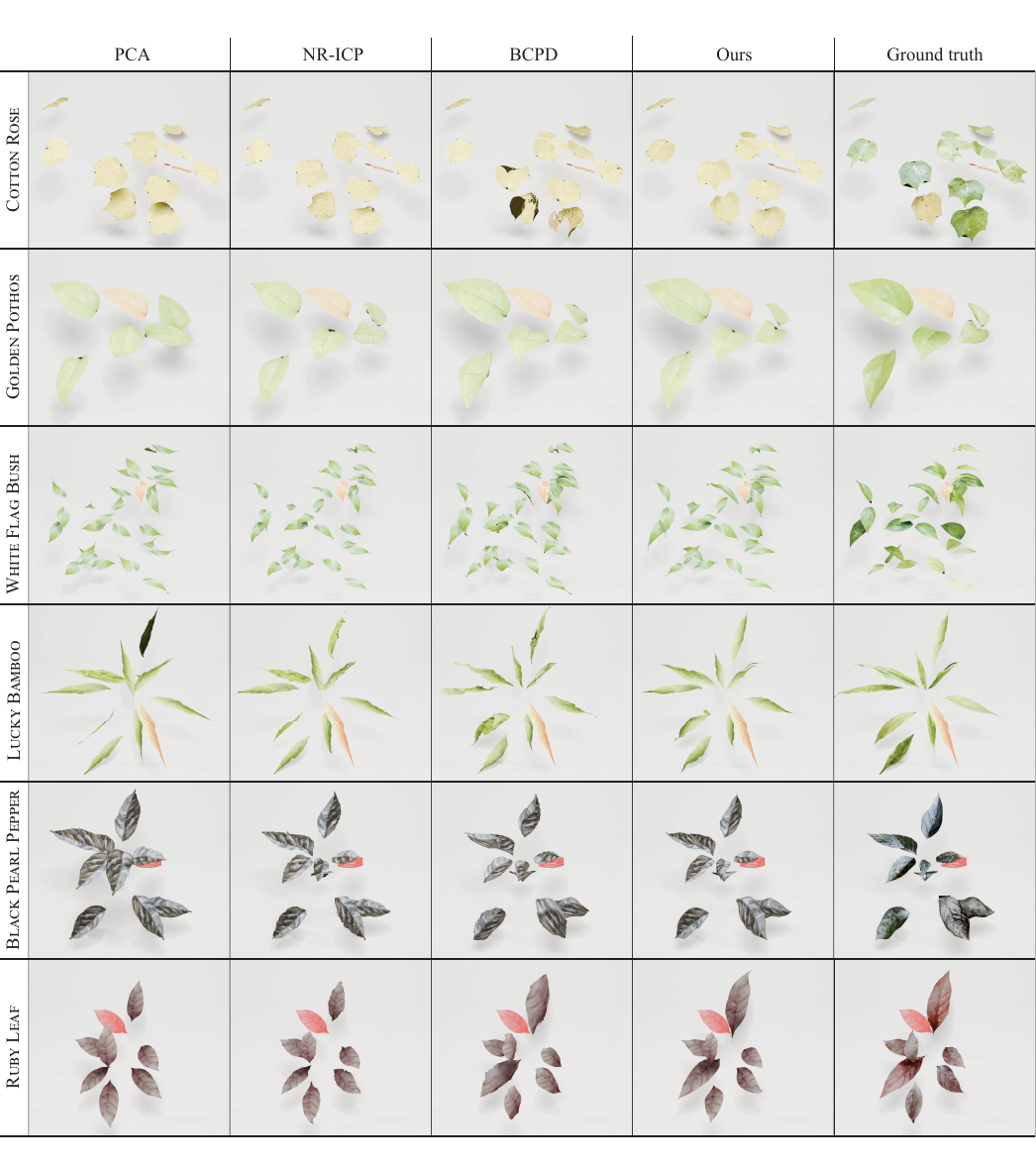}
\caption{\textbf{Deformation qualitative comparison.}
Visual results across six species comparing PCA~\cite{pearson1901pca}, NR-ICP~\cite{nricp}, BCPD~\cite{bcpd}, and Ours.
Target geometries are reconstructed from input Gaussians. While texture details naturally differ (since we warp a single template texture to match target geometry), our method exhibits the least geometric distortion and boundary misalignment compared to baselines, preserving the natural curvature of the leaves.}
  \label{fig:exp_deformation}
\end{figure*}

\paragraph*{Deformation evaluation}
We assess non-rigid alignment by deforming a single template to each target leaf and reporting Corr-\(\ell_2\) (mean nearest-neighbor error), symmetric Chamfer distance (CD)~\cite{borgefors1986chamferdistance}, and Hausdorff distance (HD)~\cite{hausdorff}. 
We compare PCA~\cite{pearson1901pca}, NR-ICP~\cite{nricp}, BCPD~\cite{bcpd}, and ours, using per-leaf meshes reconstructed from target Gaussians as ground truth. 
Regarding the non-rigid baselines, NR-ICP~\cite{nricp} extends standard ICP by solving for a smooth deformation field via a stiffness-regularized least squares objective, whereas BCPD~\cite{bcpd} treats registration as a probabilistic inference problem within a Gaussian Mixture Model, estimating coherent motion without requiring explicit point correspondences.
Qualitatively (Fig.~\ref{fig:exp_deformation}), the MLS-warped template (red) follows target contours with reduced stretching and texture distortion. Quantitatively (Tab.~\ref{tab:deformation_metrics}), our full model attains the lowest errors on all metrics.

\begin{table}[!t]
  \centering
  \begin{tabular}{
      l|
      r@{.}l   
      r@{.}l   
      r@{.}l   
  }
    \toprule
    Method
      & \multicolumn{2}{c}{Corr $\downarrow$}
      & \multicolumn{2}{c}{CD $\downarrow$}
      & \multicolumn{2}{c}{HD $\downarrow$} \\
    \midrule
    PCA~\cite{pearson1901pca}        & \dec{0}{2950}  & \dec{0}{0368}  & \dec{1}{5486} \\
    NR-ICP~\cite{nricp}              & \dec{0}{1358}  & \dec{0}{0145} & \dec{1}{0586} \\
    BCPD~\cite{bcpd}                 & \dec{0}{1186}  & \dec{0}{0061}  & \dec{0}{7554} \\
    \midrule
    Ours (w/o optim.)                & \dec{0}{1115}  & \dec{0}{0050}  & \dec{0}{6512} \\
    Ours (full)                      & \bestdec{0}{0823} & \bestdec{0}{0022} & \bestdec{0}{4669} \\
    \bottomrule
  \end{tabular}
\caption{\textbf{Deformation quantitative comparison.} Best scores are bolded. We report correspondence error (Corr), Chamfer distance (CD), and Hausdorff distance (HD) against ground-truth meshes.
Our full optimization pipeline yields the lowest error across all metrics. All values are scaled by $\times 10$ for readability.}
  \label{tab:deformation_metrics}
\end{table}

\paragraph*{Editing results}
A single template mesh is instanced across species to match diverse target leaf shapes.
In Fig.~\ref{fig:editing} (a) and (b), we showcase the result of changing the leaves of a tree into the ones from another species using MVC-based control point retargeting (see Sec.~\ref{editability}).
Leaf instances share a consistent UV layout, so edits to a texture of a template mesh propagate across all the leaves as (c) and (d) show the inpainting of the EG logo on the leaves.
In other words, changing appearance requires only swapping texture files, with no geometry edits (as demonstrated in Fig.~\ref{fig:editing} (e) and (f).
\subsection{Implementation Details}
Our prototype is primarily in Python with a lightweight GUI for fallback manual selection and live deformation previews. 
Performance-critical components (BVH construction and ray–AABB tests) are in C++: for interactive picking we build an AABB-BVH~\cite{wald2007fast} over per-Gaussian primitives, using a truncation scale of 3 to bound each primitive.
For MLS optimization and texture baking, we follow the 3DGS rasterizer~\cite{kerbl3Dgaussians} and render with an orthographic projection to produce UV textures for the extracted meshes.
Geodesic distances are computed via \texttt{geometry-central}~\cite{geometrycentral}. The  linear assignment algorithm~\cite{crouse2016implementing} mentioned in Sec.~\ref{sec:registration} is implemented by \texttt{SciPy} library.
All experiments run on an AMD Ryzen 9 9950X, an NVIDIA RTX 3090 (24\,GB), and 64\,GB RAM.

\subsection{Experimental Setup}
\label{settings}

\paragraph*{Dataset and training} We evaluate eight plants captured with an iPhone 12 mini (wide lens).
The dataset comprises six plants with roughly elliptic leaves: \textsc{Green Pepper} (9), \textsc{Rubber Tree} (14), \textsc{Golden Pothos} (5), \textsc{Black Pearl Pepper} (11), \textsc{Ruby Leaf} (7), and \textsc{White Flag Bush} (27); and two with complex morphology: \textsc{Cotton Rose} (multi-lobed, 12) and \textsc{Lucky Bamboo} (thin and elongated, 11). 
The number in parentheses denotes the leaf count.
For reconstruction, 3DGS training enables Gaussian Opacity Fields (GOF)~\cite{Yu2024GOF} to suppress floaters.
Plants are isolated from full scene reconstructions using SuperSplat’s interactive selection~\cite{playcanvas2025supersplat} (or any semantic 3D-Gaussian method~\cite{ye2024gaussian,cen2025segment}).
We will release the full dataset and code upon acceptance.

\paragraph*{Parameter settings} 
We perform global downsampling to $N_s = 8{,}192$, which empirically ensures sufficient density ($>100$ primitives) per leaf.
For efficiency, the neighbor count for graph construction is set to $N_k=512$.
Regarding segmentation thresholds, we set the triangle inequality margin $\tau=0.5$ for Eq.~\ref{eqn:tri_ineq}.
For petiole determination in Eq.~\ref{delta}, we use a geodesic band height $\delta = 0.01$ (normalized, approx. 1\,cm) to aggregate sufficient local primitives for robust estimation.
We set the petiole detection threshold $\epsilon = 0.05$ (normalized) to effectively identify the transition from the broader leaf blade to the stem.
Crucially, we apply an early protection ratio $\rho = 0.25$ to bypass naturally narrow leaf tips, preventing premature termination.
We use $K=32$ control points per leaf, which we identified as the optimal trade-off between accuracy and efficiency (see Tab.~\ref{tab:ablation_k}).
Before BPA, raw Gaussians are denoised by MLS with $\sigma=0.1$; the same $\sigma$ serves as the MLS kernel width in deformation.
Optimization runs for 200 steps with a learning rate of $7\times10^{-3}$ and Chamfer weight $\lambda=0.7$.
All hyperparameters are fixed and robust across all tested plants.

\begin{table}[b!]
  \centering
  \small
  \setlength{\tabcolsep}{6pt}
  \begin{tabular}{@{} r|ccc @{}}
    \toprule
    $K$ & Corr $\downarrow$ & CD $\downarrow$ & HD $\downarrow$ \\
    \midrule
    16             & 0.0970 & 0.0034  & 0.6030 \\
    32 (default)   & 0.0823 & 0.0021  & 0.4670 \\
    32 (w/o optim.)& 0.1115 & 0.0050 & 0.6512 \\
    64             & \textbf{0.0810} & \textbf{0.0020} & \textbf{0.4361} \\
    \bottomrule
  \end{tabular}
  \caption{\textbf{Ablation studies on number of control points $K$ and optimization.} Increasing $K$ reduces error but raises training cost. Metrics scaled by $\times 10$.}
  \label{tab:ablation_k}
\end{table}
\paragraph*{Ablation Studies}
We conduct ablation studies to investigate the influence of the control point count $K$ and to validate the necessity of the MLS optimization stage.
Increasing \(K\) adds degrees of freedom to the MLS warp, allowing it to capture finer bending or twisting; however, compute and memory costs grow linearly with \(K\). Beyond a moderate value, gains plateau, and the risk of overfitting increases.
We therefore adopt \(K{=}32\) as an accuracy–efficiency sweet spot. 
As shown in Tab.~\ref{tab:ablation_k}, increasing to \(K{=}64\) offers only marginal improvements over \(K{=}32\), whereas reducing to \(K{=}16\) or removing the optimization step (at \(K{=}32\)) significantly degrades all metrics, confirming that our optimization is essential for accurate alignment.

\section{Limitations}
\FloatBarrier
\begin{figure}[!b]
  \centering
  \includegraphics[width=84mm]{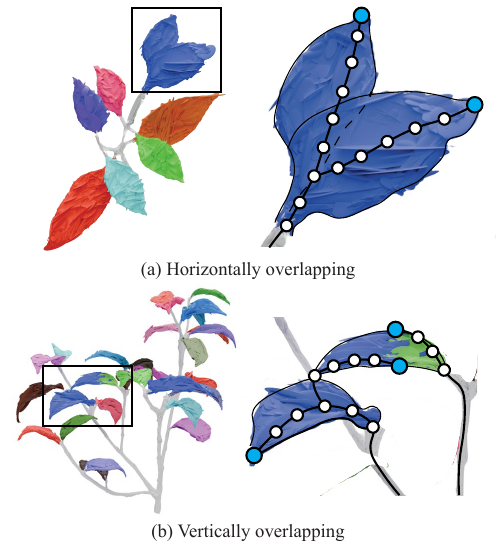}
  \caption{\textbf{Failure in overlapping cases.} The blue colored circles represent the found local minima (apexes). (a) Horizontal overlapping leads to merging paths. (b) Vertical overlapping causes the short-cut, resulting in incorrect segmentation of the upper leaf.}
  \label{fig:overlapping}
\end{figure}

Our method requires a cleaned up Gaussian model as the input and steps require manual intervention to complete the segmentation.  
Besides, our method struggles with topological ambiguity, a frequent issue in dense foliage with overlapping leaves. Since the Heat Method~\cite{Crane:2017:HMD} computes the Point Laplacian based on point proximity, overlapping leaves introduce ambiguous relations that corrupt the geodesic distance field. Fig.~\ref{fig:editing} highlights two specific failure cases, while the detected apexes are highlighted as blue colored circles. In horizontal overlapping, although the apexes are correctly detected, the grouping fails because the two paths share a large overlap in the ambiguous region cause the leafs wrongly segmented as one leave with multiple apexes. In vertical overlapping, the apex of a top leaf is spatially close to a bottom leaf, which causes the geodesic estimation around the upper leaf apex highly unreliable, as the points around are going to take the bottom leaf as a short cut to the root. 
Local maxima are found where the length of the short-cut equals the normal path (the upper two detected apexes), which leads to the incorrect segmentation at the end.


We also acknowledge that the diversity of our dataset is limited. Most plant types (except \textsc{Cotton Rose} and \textsc{Lucky Bamboo}) have simple ellipsoid-shaped leaves. However, since the ellipsoid is the most common leaf shape, we focus on this typical class in this paper. 
Handling more complex shapes remains a valuable direction for future work. For instance, in Ginkgo leaves, the local geodesic maximum often does not align with the biological apex, which leads to incorrect orientation estimation and segmentation failures which can benefit from exploiting the leaf shape symmetry.
Finally, since the leaves are deformed from a single template, the texture is shared across all leaves. This lack of unique texture variation for individual leaves is also a limitation.


\section{Conclusion}
We present LeafFit, a framework that converts plant reconstructions using dense 3D Gaussian splats to \emph{editable and instanced meshes}.
By establishing a parametric link between 3DGS and mesh assets via differentiable MLS retargeting, LeafFit reuses a single template with lightweight per-leaf controls, preserving thin-leaf detail while compressing the representation by orders of magnitude. A geodesic-guided segmentation stage yields clean instances; meshes with seamless UVs support texture and geometry edits that propagate consistently across leaves. Evaluated in a shader at runtime, the assets deliver high FPS and small memory footprints without sacrificing visual fidelity.
LeafFit integrates seamlessly with standard digital content creation tools and real-time game engines through standard mesh and texture assets, enabling artists to author, retarget, and iterate at scale.
%
%

%
Looking ahead, we find that the intrinsic symmetry of leaves is a valuable geometric cue that could be leveraged to enhance both segmentation and apex-petiole detection.
Additionally, incorporating cage-based refinement offers a promising avenue for recovering sharper and cleaner leaf silhouettes.
Finally, we anticipate that joint optimization with the Gaussian representation would benefit from a learning-based approach, where fully observed leaves serve as guidance to regularize leaves hidden by occlusion.
\section*{Acknowledgements}
This work was supported by JST SPRING, Grant Number JPMJSP2108. We would like to thank Haato Watanabe and Yudi Wu for their assistance with data collection, as well as the anonymous reviewers for their constructive comments which helped improve the manuscript.
\bibliographystyle{eg-alpha-doi} 
\bibliography{egbibsample}       

@article{Yu2024GOF,
  author    = {Yu, Zehao and Sattler, Torsten and Geiger, Andreas},
  title     = {Gaussian Opacity Fields: Efficient Adaptive Surface Reconstruction in Unbounded Scenes},
  journal   = {ACM Transactions on Graphics},
  year      = {2024},
}

@article{guedon2023sugar,
  title={SuGaR: Surface-Aligned Gaussian Splatting for Efficient 3D Mesh Reconstruction and High-Quality Mesh Rendering},
  author={Gu{\'e}don, Antoine and Lepetit, Vincent},
  journal={CVPR},
  year={2024}
}

@article{yang2025neuraleaf,
  author    = {Yang, Yang and Dongni, Mao and Hiroaki, Santo and Yasuyuki, Matsushita and Fumio, Okura},
  title     = {NeuraLeaf: Neural Parametric Leaf Modelswith Shape and Deformation Disentanglement},
  journal   = {ICCV},
  year      = {2025},
}

@inproceedings{Huang2DGS2024,
    title={2D Gaussian Splatting for Geometrically Accurate Radiance Fields},
    author={Huang, Binbin and Yu, Zehao and Chen, Anpei and Geiger, Andreas and Gao, Shenghua},
    publisher = {Association for Computing Machinery},
    booktitle = {SIGGRAPH 2024 Conference Papers},
    year      = {2024},
    doi       = {10.1145/3641519.3657428}
}

@Article{kerbl3Dgaussians,
      author       = {Kerbl, Bernhard and Kopanas, Georgios and Leimk{\"u}hler, Thomas and Drettakis, George},
      title        = {3D Gaussian Splatting for Real-Time Radiance Field Rendering},
      journal      = {ACM Transactions on Graphics},
      number       = {4},
      volume       = {42},
      month        = {July},
      year         = {2023},
      url          = {https://repo-sam.inria.fr/fungraph/3d-gaussian-splatting/}
}

@software{maya,
  author = {{Autodesk, INC.}},
  title = {Maya},
  url = {https:/ autodesk.com/maya},
  version = {2019},
  date = {2019-01-15},
}

@Manual{blender,
   title = {Blender - a 3D modelling and rendering package},
   author = {Blender Online Community},
   organization = {Blender Foundation},
   address = {Stichting Blender Foundation, Amsterdam},
   year = {2018},
   url = {http://www.blender.org},
 }

@inproceedings{mildenhall2020nerf,
  title={NeRF: Representing Scenes as Neural Radiance Fields for View Synthesis},
  author={Ben Mildenhall and Pratul P. Srinivasan and Matthew Tancik and Jonathan T. Barron and Ravi Ramamoorthi and Ren Ng},
  year={2020},
  booktitle={ECCV},
}

@misc{li2024procgs,
      title={Proc-GS: Procedural Building Generation for City Assembly with 3D Gaussians}, 
      author={Yixuan Li and Xingjian Ran and Linning Xu and Tao Lu and Mulin Yu and Zhenzhi Wang and Yuanbo Xiangli and Dahua Lin and Bo Dai},
      year={2024},
      eprint={2412.07660},
      archivePrefix={arXiv},
      primaryClass={cs.CV},
      url={https://arxiv.org/abs/2412.07660}, 
}

@inproceedings{huang2024DRK,
  title={Deformable Radial Kernel Splatting},
  author={Huang, Yi-Hua and Lin, Ming-Xian and Sun, Yang-Tian and Yang, Ziyi and Lyu, Xiaoyang and Cao, Yan-Pei and Qi, Xiaojuan},
  booktitle={Proceedings of the Computer Vision and Pattern Recognition Conference},
  pages={21513--21523},
  year={2025}
}

@inproceedings{schoenberger2016sfm,
    author={Sch\"{o}nberger, Johannes Lutz and Frahm, Jan-Michael},
    title={Structure-from-Motion Revisited},
    booktitle={Conference on Computer Vision and Pattern Recognition (CVPR)},
    year={2016},
}

@inproceedings{schoenberger2016mvs,
    author={Sch\"{o}nberger, Johannes Lutz and Zheng, Enliang and Pollefeys, Marc and Frahm, Jan-Michael},
    title={Pixelwise View Selection for Unstructured Multi-View Stereo},
    booktitle={European Conference on Computer Vision (ECCV)},
    year={2016},
}

@inproceedings{nguyen2016comparison,
  title={Comparison of structure-from-motion and stereo vision techniques for full in-field 3d reconstruction and phenotyping of plants: An investigation in sunflower},
  author={Nguyen, Thuy Tuong and Slaughter, David C and Townsley, Brad and Carriedo, Leonela and NN, Julin and Sinha, Neelima},
  booktitle={2016 ASABE annual international meeting},
  pages={1},
  year={2016},
  organization={American Society of Agricultural and Biological Engineers}
}

@article{andujar2018three,
  title={Three-dimensional modeling of weed plants using low-cost photogrammetry},
  author={And{\'u}jar, Dionisio and Calle, Mikel and Fern{\'a}ndez-Quintanilla, C{\'e}sar and Ribeiro, {\'A}ngela and Dorado, Jos{\'e}},
  journal={Sensors},
  volume={18},
  number={4},
  pages={1077},
  year={2018},
  publisher={MDPI}
}

@article{lu2020reconstruction,
  title={Reconstruction method and optimum range of camera-shooting angle for 3D plant modeling using a multi-camera photography system},
  author={Lu, Xingtong and Ono, Eiichi and Lu, Shan and Zhang, Yu and Teng, Poching and Aono, Mitsuko and Shimizu, Yo and Hosoi, Fumiki and Omasa, Kenji},
  journal={Plant Methods},
  volume={16},
  number={1},
  pages={118},
  year={2020},
  publisher={Springer}
}

@article{pradal2009plantgl,
  title={PlantGL: a Python-based geometric library for 3D plant modelling at different scales},
  author={Pradal, Christophe and Boudon, Fr{\'e}d{\'e}ric and Nouguier, Christophe and Chopard, J{\'e}r{\^o}me and Godin, Christophe},
  journal={Graphical models},
  volume={71},
  number={1},
  pages={1--21},
  year={2009},
  publisher={Elsevier}
}

@article{mueller2022instant,
    author = {Thomas M\"uller and Alex Evans and Christoph Schied and Alexander Keller},
    title = {Instant Neural Graphics Primitives with a Multiresolution Hash Encoding},
    journal = {ACM Trans. Graph.},
    issue_date = {July 2022},
    volume = {41},
    number = {4},
    month = jul,
    year = {2022},
    pages = {102:1--102:15},
    articleno = {102},
    numpages = {15},
    url = {https://doi.org/10.1145/3528223.3530127},
    doi = {10.1145/3528223.3530127},
    publisher = {ACM},
    address = {New York, NY, USA},
}

@article{hu2024high,
  title={High-fidelity 3D reconstruction of plants using Neural Radiance Fields},
  author={Hu, Kewei and Ying, Wei and Pan, Yaoqiang and Kang, Hanwen and Chen, Chao},
  journal={Computers and Electronics in Agriculture},
  volume={220},
  pages={108848},
  year={2024},
  publisher={Elsevier}
}

@inproceedings{wang2024gaussianeditor,
  title={Gaussianeditor: Editing 3d gaussians delicately with text instructions},
  author={Wang, Junjie and Fang, Jiemin and Zhang, Xiaopeng and Xie, Lingxi and Tian, Qi},
  booktitle={Proceedings of the IEEE/CVF conference on computer vision and pattern recognition},
  pages={20902--20911},
  year={2024}
}

@article{paproki2012novel,
  title={A novel mesh processing based technique for 3D plant analysis},
  author={Paproki, Anthony and Sirault, Xavier and Berry, Scott and Furbank, Robert and Fripp, Jurgen},
  journal={BMC plant biology},
  volume={12},
  number={1},
  pages={63},
  year={2012},
  publisher={Springer}
}

@article{paulus2019measuring,
  title={Measuring crops in 3D: using geometry for plant phenotyping},
  author={Paulus, Stefan},
  journal={Plant methods},
  volume={15},
  number={1},
  pages={103},
  year={2019},
  publisher={Springer}
}

@article{okura20223d,
  title={3D modeling and reconstruction of plants and trees: A cross-cutting review across computer graphics, vision, and plant phenotyping},
  author={Okura, Fumio},
  journal={Breeding Science},
  volume={72},
  number={1},
  pages={31--47},
  year={2022},
  publisher={Japanese Society of Breeding}
}

@misc{ORCASpeedTree,
   title = {SpeedTree, Open Research Content Archive (ORCA)},
   author = {SpeedTree},
   year = {2017},
   month = {July},
   url = {http://developer.nvidia.com/orca/speedtree}
}

@inproceedings{haque2023instruct,
  title={Instruct-nerf2nerf: Editing 3d scenes with instructions},
  author={Haque, Ayaan and Tancik, Matthew and Efros, Alexei A and Holynski, Aleksander and Kanazawa, Angjoo},
  booktitle={Proceedings of the IEEE/CVF international conference on computer vision},
  pages={19740--19750},
  year={2023}
}

@inproceedings{li2025plantsurvey,
  title={A survey on 3D reconstruction techniques in plant phenotyping: from classical methods to neural radiance fields (NeRF), 3D Gaussian splatting (3DGS), and beyond},
  author={Li, Jiajia and Qi, Xinda and Nabaei, Seyed H and Chen, Dong and Zhang, Xin and Li, Zhaojian},
  booktitle={Autonomous Air and Ground Sensing Systems for Agricultural Optimization and Phenotyping X},
  volume={13475},
  pages={134750B},
  year={2025},
  organization={SPIE}
}

@inproceedings{fruitnerf2024,
  author     = {Lukas Meyer and Andreas Gilson and Ute Schmidt and Marc Stamminger},
 title      = {FruitNeRF: A Unified Neural Radiance Field based Fruit Counting Framework},
 booktitle  = {IROS},
 year       = {2024},
 url        = {https://meyerls.github.io/fruit_nerf}
}

@incollection{ojo2024splanting,
  title={Splanting: 3d plant capture with gaussian splatting},
  author={Ojo, Tommy and La, Thai and Morton, Andrew and Stavness, Ian},
  booktitle={SIGGRAPH Asia 2024 Technical Communications},
  pages={1--4},
  year={2024}
}

@article{shen2025plantgaussian,
  title={PlantGaussian: exploring 3d gaussian splatting for cross-time, cross-scene, and realistic 3d plant visualization and beyond},
  author={Shen, Peng and Jing, Xueyao and Deng, Wenzhe and Jia, Hanyue and Wu, Tingting},
  journal={The Crop Journal},
  volume={13},
  number={2},
  pages={607--618},
  year={2025},
  publisher={Elsevier}
}

@inproceedings{ijiri2006sketch,
  title={The sketch l-system: Global control of tree modeling using free-form strokes},
  author={Ijiri, Takashi and Owada, Shigeru and Igarashi, Takeo},
  booktitle={International symposium on smart graphics},
  pages={138--146},
  year={2006},
  organization={Springer}
}

@article{bradley2013image,
  title={Image-based reconstruction and synthesis of dense foliage},
  author={Bradley, Derek and Nowrouzezahrai, Derek and Beardsley, Paul},
  journal={ACM Transactions on Graphics (TOG)},
  volume={32},
  number={4},
  pages={1--10},
  year={2013},
  publisher={ACM New York, NY, USA}
}

@inproceedings{chaurasia2017editable,
  title={Editable parametric dense foliage from 3D capture},
  author={Chaurasia, Gaurav and Beardsley, Paul},
  booktitle={Proceedings of the IEEE International Conference on Computer Vision},
  pages={5305--5314},
  year={2017}
}

@article{guo2018realistic,
  title={Realistic procedural plant modeling from multiple view images},
  author={Guo, Jianwei and Xu, Shibiao and Yan, Dong-Ming and Cheng, Zhanglin and Jaeger, Marc and Zhang, Xiaopeng},
  journal={IEEE transactions on visualization and computer graphics},
  volume={26},
  number={2},
  pages={1372--1384},
  year={2018},
  publisher={IEEE}
}

@inproceedings{stava2014inverse,
  title={Inverse procedural modelling of trees},
  author={Stava, Ondrej and Pirk, S{\"o}ren and Kratt, Julian and Chen, Baoquan and M{\v{e}}ch, Radom{\'\i}r and Deussen, Oliver and Benes, Bedrich},
  booktitle={Computer Graphics Forum},
  volume={33},
  number={6},
  pages={118--131},
  year={2014},
  organization={Wiley Online Library}
}

@article{talton2011metropolis,
  title={Metropolis procedural modeling.},
  author={Talton, Jerry O and Lou, Yu and Lesser, Steve and Duke, Jared and Mech, Radom{\'\i}r and Koltun, Vladlen},
  journal={ACM Trans. Graph.},
  volume={30},
  number={2},
  pages={11--1},
  year={2011}
}

@software{playcanvas2025supersplat,
  author       = {{PlayCanvas} and {Snap Inc.}},
  title        = {SuperSplat},
  year         = {2025},
  url          = {https://github.com/playcanvas/supersplat},
  note         = {[Computer software]},
}

@inproceedings{jaingaussiancut,
  title={GaussianCut: Interactive segmentation via graph cut for 3D Gaussian Splatting},
  author={Jain, Umangi and Mirzaei, Ashkan and Gilitschenski, Igor},
  booktitle={The Thirty-eighth Annual Conference on Neural Information Processing Systems},
  year={2024}
}

@inproceedings{xiao2025localized,
  title={Localized gaussian splatting editing with contextual awareness},
  author={Xiao, Hanyuan and Chen, Yingshu and Huang, Huajian and Xiong, Haolin and Yang, Jing and Prasad, Pratusha and Zhao, Yajie},
  booktitle={2025 IEEE/CVF Winter Conference on Applications of Computer Vision (WACV)},
  pages={5207--5217},
  year={2025},
  organization={IEEE}
}

@article{yan20243dsceneeditor,
  title={3dsceneeditor: Controllable 3d scene editing with gaussian splatting},
  author={Yan, Ziyang and Li, Lei and Shao, Yihua and Chen, Siyu and Wu, Zongkai and Hwang, Jenq-Neng and Zhao, Hao and Remondino, Fabio},
  journal={arXiv preprint arXiv:2412.01583},
  year={2024}
}

@article{Crane:2017:HMD,
 author = {Crane, Keenan and Weischedel, Clarisse and Wardetzky, Max},
 title = {The Heat Method for Distance Computation},
 journal = {Commun. ACM},
 issue_date = {November 2017},
 volume = {60},
 number = {11},
 month = oct,
 year = {2017},
 issn = {0001-0782},
 pages = {90--99},
 numpages = {10},
 url = {http://doi.acm.org/10.1145/3131280},
 doi = {10.1145/3131280},
 acmid = {3131280},
 publisher = {ACM},
 address = {New York, NY, USA},
}

@article{jiang2024vr-gs,
      title={VR-GS: A Physical Dynamics-Aware Interactive Gaussian Splatting System in Virtual Reality},
      author={Jiang, Ying and Yu, Chang and Xie, Tianyi and Li, Xuan and Feng, Yutao and Wang, Huamin and Li, Minchen and Lau, Henry and Gao, Feng and Yang, Yin and Jiang, Chenfanfu},
      journal={arXiv preprint arXiv:2401.16663},
      year={2024},
}

@article{xie2024sketch,
  title={Sketch-guided cage-based 3d gaussian splatting deformation},
  author={Xie, Tianhao and Aigerman, Noam and Belilovsky, Eugene and Popa, Tiberiu},
  journal={arXiv preprint arXiv:2411.12168},
  year={2024}
}

@article{huang2024gsdeformer,
  title={GSDeformer: Direct, Real-time and Extensible Cage-based Deformation for 3D Gaussian Splatting},
  author={Huang, Jiajun and Xu, Shuolin and Yu, Hongchuan and Lee, Tong-Yee},
  journal={arXiv preprint arXiv:2405.15491},
  year={2024}
}

@inproceedings{chang2024gaussreg,
  title={Gaussreg: Fast 3d registration with gaussian splatting},
  author={Chang, Jiahao and Xu, Yinglin and Li, Yihao and Chen, Yuantao and Feng, Wensen and Han, Xiaoguang},
  booktitle={European Conference on Computer Vision},
  pages={407--423},
  year={2024},
  organization={Springer}
}

@inproceedings{segal2009generalized,
  title={Generalized-icp.},
  author={Segal, Aleksandr and Haehnel, Dirk and Thrun, Sebastian},
  booktitle={Robotics: science and systems},
  volume={2},
  number={4},
  pages={435},
  year={2009},
  organization={Seattle, WA}
}

@article{zhu20073d,
  title={3D deformation using moving least squares},
  author={Zhu, Yuanchen and Gortler, Steven},
  year={2007}
}

@incollection{schaefer2006image,
  title={Image deformation using moving least squares},
  author={Schaefer, Scott and McPhail, Travis and Warren, Joe},
  booktitle={ACM SIGGRAPH 2006 Papers},
  pages={533--540},
  year={2006}
}

@incollection{lorensen1998marching,
  title={Marching cubes: A high resolution 3D surface construction algorithm},
  author={Lorensen, William E and Cline, Harvey E},
  booktitle={Seminal graphics: pioneering efforts that shaped the field},
  pages={347--353},
  year={1998}
}

@inproceedings{park2019deepsdf,
  title={Deepsdf: Learning continuous signed distance functions for shape representation},
  author={Park, Jeong Joon and Florence, Peter and Straub, Julian and Newcombe, Richard and Lovegrove, Steven},
  booktitle={Proceedings of the IEEE/CVF conference on computer vision and pattern recognition},
  pages={165--174},
  year={2019}
}

@article{bernardini2002ball,
  title={The ball-pivoting algorithm for surface reconstruction},
  author={Bernardini, Fausto and Mittleman, Joshua and Rushmeier, Holly and Silva, Cl{\'a}udio and Taubin, Gabriel},
  journal={IEEE transactions on visualization and computer graphics},
  volume={5},
  number={4},
  pages={349--359},
  year={2002},
  publisher={IEEE}
}

@inproceedings{pandey2025painting,
  title={Painting with 3D Gaussian Splat Brushes},
  author={Pandey, Karran and Hu, Anita and Tsang, Clement Fuji and Perel, Or and Singh, Karan and Shugrina, Maria},
  booktitle={Proceedings of the Special Interest Group on Computer Graphics and Interactive Techniques Conference Conference Papers},
  pages={1--10},
  year={2025}
}

@inproceedings{wald2007fast,
  title={On fast construction of SAH-based bounding volume hierarchies},
  author={Wald, Ingo},
  booktitle={2007 IEEE Symposium on Interactive Ray Tracing},
  pages={33--40},
  year={2007},
  organization={IEEE}
}

@article{friedman1977kdtree,
  title={An algorithm for finding best matches in logarithmic expected time},
  author={Friedman, Jerome H and Bentley, Jon Louis and Finkel, Raphael Ari},
  journal={ACM Transactions on Mathematical Software (TOMS)},
  volume={3},
  number={3},
  pages={209--226},
  year={1977},
  publisher={ACM New York, NY, USA}
}

@article{pearson1901pca,
  title={LIII. On lines and planes of closest fit to systems of points in space},
  author={Pearson, Karl},
  journal={The London, Edinburgh, and Dublin philosophical magazine and journal of science},
  volume={2},
  number={11},
  pages={559--572},
  year={1901},
  publisher={Taylor \& Francis}
}

@article{crouse2016implementing,
  title={On implementing 2D rectangular assignment algorithms},
  author={Crouse, David F},
  journal={IEEE Transactions on Aerospace and Electronic Systems},
  volume={52},
  number={4},
  pages={1679--1696},
  year={2016},
  publisher={IEEE}
}

@inproceedings{wang2013method,
  title={A method of realistic leaves modeling based on point cloud},
  author={Wang, Yinghui and Hao, Wen and Wang, Gang and Ning, Xiaojuan and Tang, Jing and Shi, Zhenghao and Wang, Ningna and Zhao, Minghua},
  booktitle={Proceedings of the 12th ACM SIGGRAPH International Conference on Virtual-Reality Continuum and Its Applications in Industry},
  pages={123--130},
  year={2013}
}

@article{violante2025splatandreplace,
  title={Splat and Replace: 3D Reconstruction with Repetitive Elements},
  author={Violante, Nicolás and Meuleman, Andreas and Gauthier, Alban and Durand, Fredo and Groueix, Thibault and Drettakis, George},
  journal={SIGGRAPH Conference Papers},
  year={2025}
}

@inproceedings{gsicpslam2024,
author = {Ha, Seongbo and Yeon, Jiung and Yu, Hyeonwoo},
title = {RGBD GS-ICP SLAM},
year = {2024},
isbn = {978-3-031-72763-4},
publisher = {Springer-Verlag},
address = {Berlin, Heidelberg},
url = {https://doi.org/10.1007/978-3-031-72764-1_11},
doi = {10.1007/978-3-031-72764-1_11},
booktitle = {Computer Vision – ECCV 2024: 18th European Conference, Milan, Italy, September 29–October 4, 2024, Proceedings, Part XXXVI},
pages = {180–197},
numpages = {18},
location = {Milan, Italy}
}

@article{geometrycentral,
  title={GeometryCentral: A modern C++ library of data structures and algorithms for geometry processing},
  author={Nicholas Sharp and Keenan Crane and others},
  howpublished="\url{https://geometry-central.net/}",
  year={2019}
}

@article{borgefors1986chamferdistance,
  title={Distance transformations in digital images},
  author={Borgefors, Gunilla},
  journal={Computer vision, graphics, and image processing},
  volume={34},
  number={3},
  pages={344--371},
  year={1986},
  publisher={Elsevier}
}

@INPROCEEDINGS{nricp,
  author={Amberg, Brian and Romdhani, Sami and Vetter, Thomas},
  booktitle={2007 IEEE Conference on Computer Vision and Pattern Recognition}, 
  title={Optimal Step Nonrigid ICP Algorithms for Surface Registration}, 
  year={2007},
  volume={},
  number={},
  pages={1-8},
  doi={10.1109/CVPR.2007.383165}}

@ARTICLE{bcpd,
  author={Hirose, Osamu},
  journal={IEEE Transactions on Pattern Analysis and Machine Intelligence}, 
  title={Geodesic-Based Bayesian Coherent Point Drift}, 
  year={2023},
  volume={45},
  number={5},
  pages={5816-5832},
  doi={10.1109/TPAMI.2022.3214191}}

@inproceedings{wu2024ptv3,
    title={Point Transformer V3: Simpler, Faster, Stronger},
    author={Wu, Xiaoyang and Jiang, Li and Wang, Peng-Shuai and Liu, Zhijian and Liu, Xihui and Qiao, Yu and Ouyang, Wanli and He, Tong and Zhao, Hengshuang},
    booktitle={CVPR},
    year={2024}
}

@ARTICLE{hausdorff,
  author={Huttenlocher, D.P. and Klanderman, G.A. and Rucklidge, W.J.},
  journal={IEEE Transactions on Pattern Analysis and Machine Intelligence}, 
  title={Comparing images using the Hausdorff distance}, 
  year={1993},
  volume={15},
  number={9},
  pages={850-863},
  doi={10.1109/34.232073}}

@article{mvc2003,
author = {Floater, Michael S.},
title = {Mean value coordinates},
year = {2003},
issue_date = {March 2003},
publisher = {Elsevier Science Publishers B. V.},
address = {NLD},
volume = {20},
number = {1},
issn = {0167-8396},
journal = {Comput. Aided Geom. Des.},
month = mar,
pages = {19–27},
numpages = {9}
}

@inproceedings{holes_in_mesh,
booktitle = {Eurographics Symposium on Geometry Processing},
editor = {Leif Kobbelt and Peter Schroeder and Hugues Hoppe},
title = {{Filling Holes in Meshes}},
author = {Liepa, Peter},
year = {2003},
publisher = {The Eurographics Association},
ISSN = {1727-8384},
ISBN = {3-905673-06-1},
DOI = {/10.2312/SGP/SGP03/200-206}
}

@article{mertouglu2024planest,
  title={PLANesT-3D: A new annotated dataset for segmentation of 3D plant point clouds},
  author={Merto{\u{g}}lu, Kerem and {\c{S}}alk, Yusuf and Sar{\i}kaya, Server Karahan and Turgut, Kaya and Evreneso{\u{g}}lu, Yasemin and {\c{C}}evikalp, Hakan and Gerek, {\"O}mer Nezih and Duta{\u{g}}ac{\i}, Helin and Rousseau, David},
  journal={arXiv preprint arXiv:2407.21150},
  year={2024}
}

@inproceedings{ye2024gaussian,
  title={Gaussian grouping: Segment and edit anything in 3d scenes},
  author={Ye, Mingqiao and Danelljan, Martin and Yu, Fisher and Ke, Lei},
  booktitle={European conference on computer vision},
  pages={162--179},
  year={2024},
  organization={Springer}
}

@inproceedings{cen2025segment,
  title={Segment any 3d gaussians},
  author={Cen, Jiazhong and Fang, Jiemin and Yang, Chen and Xie, Lingxi and Zhang, Xiaopeng and Shen, Wei and Tian, Qi},
  booktitle={Proceedings of the AAAI Conference on Artificial Intelligence},
  volume={39},
  number={2},
  pages={1971--1979},
  year={2025}
}

@article{amenta2004defining,
author = {Amenta, Nina and Kil, Yong Joo},
title = {Defining point-set surfaces},
year = {2004},
issue_date = {August 2004},
publisher = {Association for Computing Machinery},
address = {New York, NY, USA},
volume = {23},
number = {3},
issn = {0730-0301},
url = {https://doi.org/10.1145/1015706.1015713},
doi = {10.1145/1015706.1015713},
journal = {ACM Trans. Graph.},
month = aug,
pages = {264–270},
numpages = {7}
}

@article{Soneira1988Leonardo,
  title        = {Leonardo's rule, self-similarity and random walks in trees},
  author       = {Soneira, Jose},
  journal      = {Physica A: Statistical Mechanics and its Applications},
  volume       = {149},
  number       = {3-4},
  pages        = {641--652},
  year         = {1988},
  publisher    = {Elsevier},
  doi          = {10.1016/0378-4371(88)90293-9}
}

@inproceedings{kirillov2019panoptic,
  title={Panoptic segmentation},
  author={Kirillov, Alexander and He, Kaiming and Girshick, Ross and Rother, Carsten and Doll{\'a}r, Piotr},
  booktitle={Proceedings of the IEEE/CVF conference on computer vision and pattern recognition},
  pages={9404--9413},
  year={2019}
}

@article{bourquat2022hierarchical,
  title={Hierarchical mesh-to-points as-rigid-as-possible registration},
  author={Bourquat, Pierre and Coeurjolly, David and Damiand, Guillaume and Dupont, Florent},
  journal={Computers \& Graphics},
  volume={102},
  pages={320--328},
  year={2022},
  publisher={Elsevier}
}

@article{SONG2025296,
title = {Comprehensive review on 3D point cloud segmentation in plants},
journal = {Artificial Intelligence in Agriculture},
volume = {15},
number = {2},
pages = {296-315},
year = {2025},
issn = {2589-7217},
doi = {https://doi.org/10.1016/j.aiia.2025.01.006},
url = {https://www.sciencedirect.com/science/article/pii/S2589721725000066},
author = {Hongli Song and Weiliang Wen and Sheng Wu and Xinyu Guo}
}

@book{prusinkiewicz2012algorithmic,
  title={The algorithmic beauty of plants},
  author={Prusinkiewicz, Przemyslaw and Lindenmayer, Aristid},
  year={2012},
  publisher={Springer Science \& Business Media}
}

@article{lee2023latent,
  title={Latent l-systems: Transformer-based tree generator},
  author={Lee, Jae Joong and Li, Bosheng and Benes, Bedrich},
  journal={ACM Transactions on Graphics},
  volume={43},
  number={1},
  pages={1--16},
  year={2023},
  publisher={ACM New York, NY}
}

@article{zhou2025treestructor,
  title={TreeStructor: Forest Reconstruction with Neural Ranking},
  author={Zhou, Xiaochen and Li, Bosheng and Benes, Bedrich and Habib, Ayman and Fei, Songlin and Shao, Jinyuan and Pirk, S{\"o}ren},
  journal={IEEE Transactions on Geoscience and Remote Sensing},
  year={2025},
  publisher={IEEE}
}


\end{document}